\documentclass[twocolumn]{aastex63}
\usepackage{amsmath}
\usepackage{multirow}
\usepackage{booktabs}
\let\tablenum\relax
\usepackage{siunitx}
\newcommand{\be}{\begin{equation}}
\newcommand{\ee}{\end{equation}}
\newcommand{\bea}{\begin{eqnarray}}
\renewcommand{\thefootnote}{\fnsymbol{footnote}}
\newcommand{\eea}{\end{eqnarray}}

\shorttitle{A Six Planet Resonance Chain in K2-138?}
\shortauthors{M. Cerioni \& C. Beaug\'e}
\graphicspath{{./}{}}
\begin{document}
\title{A Six Planet Resonance Chain in K2-138?}
\correspondingauthor{Matías Cerioni}
\email{matias.cerioni@unc.edu.ar}
\author{M. Cerioni}
\author{C. Beaugé}
\affiliation{Instituto de Astronomía Teórica y Experimental (IATE), \\
Observatorio Astronómico, Universidad Nacional de Córdoba, \\
Laprida 854, (X5000BGR) Córdoba, Argentina}

\begin{abstract}
The K2-138 system hosts six planets and presents an 
interesting case study due to its distinctive dynamical structure. Its five 
inner planets are near a chain of 3/2 two-body mean-motion resonances, while the 
outermost body (planet {\it g}) is significantly detached, having a mean-motion 
ratio of $n_f/n_g \sim 3.3$ with its closest neighbor. We show that the orbit of $m_g$ is actually consistent with the first-order 
three-planet resonance (3P-MMR) characterized by the relation $2n_e - 4n_f 
+ 3n_g = 0$ and is the first time a pure 
first-order 3P-MMR is found in a multi-planet system and tied to its current 
dynamical structure. Adequate values for the masses allow to trace the dynamical history of the 
system from an initial capture in a 6-planet chain (with $n_f/n_g$ in a 3/1 resonance), up to its current 
configuration due to tidal interactions over the age of the star. 
The increase in resonance offset with semi-major axis, as well as its large 
value for $n_f/n_g$ can be explained by the slopes of the pure three-planet 
resonances in the mean-motion ratio plane. The triplets slide outward over these 
curves when the innermost pair is pulled apart by tidal effects, in a 
\textit{pantograph-}like manner. The capture into the 3P-MMR is found to be surprisingly robust given 
similar masses for $m_g$ and $m_f$, and it is possible that the same effect 
may also be found in other compact planetary systems.
\end{abstract}


\section{Introduction} \label{sec:introduction}

One of the things we learnt from the \textit{Kepler Space Telescope} is about the 
significant lack of 2-planet mean-motion resonances (2P-MMRs) among low-mass 
planets ($< 20 m_\oplus$) when compared to their high-mass counterparts ($>100 
m_\oplus$) \citep{Lissauer.etal.2011,Winn.Fabrycky.2015}. Nevertheless, we 
still find rich resonant dynamics in the former domain.

On one hand, it has been recently argued that, even when lacking in 
\textit{pair}-wise commensurabilities, their distribution of orbital periods 
shows a correlation with the web of 3-planet mean-motion resonances (3P-MMRs) 
\citep{Cerioni.etal.2022}, suggesting a more complex resonant evolution than 
previously thought. On the other hand, planetary systems showing a high number 
of MMRs among its constituents are predominantly low-mass (e.g Kepler-60, 
Kepler-80, Kepler-223, TRAPPIST-1, TOI-178, K2-138, etc.), although some 
examples are also found in systems harboring massive planets (e.g. GJ876 and 
HR8799). These multi-resonant systems appear as "resonance chains" with multiple 
links of adjacent 2P-MMRs and (consequently or not) zero-order 3P-MMRs.

We do not expect these commensurabilities to be the outcome of \textit{in-situ} 
formation with fortunate initial conditions. That scheme falls short because it 
fails to account for the meaningful gravitational interactions between 
planetesimals and a gaseous protoplanetary disk, mainly, disk-induced migration.
This mechanism is characterised by a smooth decrease in semi-major axis, 
forcing a slow inward spiraling of the bodies towards the star.

\begin{figure}[t]
\centering
\includegraphics*[scale=.79]{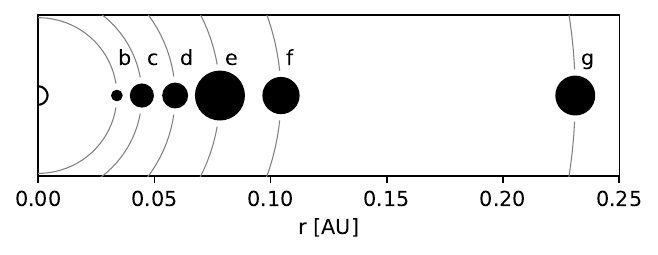}
\caption{Sketch of K2-138. The white circle on the left corresponds to the 
central star. The solid lines correspond to their semi-major axes as listed in 
Table \ref{tab:planets}. Black dot sizes are proportional to the measured 
radii.}
\label{fig:sketch}
\end{figure}

Depending on their relative rates, migration can lead two planets into a 
resonance. 
Although in the case of divergent migration (where planet separation increases) planets are unlikely to become trapped in resonance \citep[e.g,][]{Henrard.Lemaitre.1983}, the case of convergent migration (where planet separation decreases) is different. 
Convergent migration can lead to permanent capture into resonance \citep{Goldreich.Tremaine.1980, Lee.Peale.2002}, and is believed to be the cause of most observed MMRs, such as that of Neptune and Pluto \citep{Malhotra.1991}, the satellites of Jupiter and Saturn, as well as several large-mass exoplanetary systems \cite[e.g.,][]{Lee.Peale.2002, Beauge.etal.2003, Ramos.etal.2017}.

It is believed that resonance capture in stable solutions of first-order MMRs 
is highly probable provided sufficiently slow smooth differential migration and 
low initial eccentricities \citep[see][]{Henrard.Lemaitre.1983, 
Beauge.etal.2006, Batygin.2015}. Resonance chains are therefore expected to be a 
frequent outcome given any system of multiple migrating low-mass planetesimals, 
although many may succumb under instabilities induced by the other planets and 
disrupt after disk dispersal 
\citep[e.g.,][]{Thommes.etal.2008,Lega.etal.2013,Izidoro.etal.2017}.

One resonant system that recently caught our attention is K2-138, the first K2 
system discovered by citizen scientists through the Exoplanet Explorers program 
on the Zooniverse platform \citep{Christiansen.etal.2018}.

K2-138 is a K-Dwarf type star with mass $m_0 = 0.93 \pm 0.02$ $m_\odot$ and 
radius $R_0 = 0.86 \pm 0.03$ $R_\odot$, hosting one super-Earth and 
five sub-Neptunes \citep{Christiansen.etal.2018, Lopez.etal.2019, 
Hardegree-ullman.etal.2021}. A sketch of the system's architecture can be seen 
in Figure \ref{fig:sketch}. In addition to the Kepler lightcurves, HARPS radial 
velocity measurements and Spitzer lightcurves have allowed to simultaneously 
constrain planetary orbits, radii and (in some cases) masses. The 
complete list of K2 and HARPS derived planetary parameters are presented in 
Table \ref{tab:planets}.

\begin{deluxetable*}{cDDDDD}
\tablecaption{K2-138 planetary parameters as derived and presented in \cite{Lopez.etal.2019}.}
\label{tab:planets}
\tablewidth{0pt}
\tablehead{
\colhead{Planets} & \multicolumn2c{P} & \multicolumn2c{a} & \multicolumn2c{e} & \multicolumn2c{R} & \multicolumn2c{\textbf{m}} \\
\colhead{ } & \multicolumn2c{(days)} & \multicolumn2c{(AU)} & \multicolumn2c{ } 
 & \multicolumn2c{($R_\oplus$)} & \multicolumn2c{($m_\oplus$)}
}
\decimals
\startdata
 b & 2.35309  $^{+0.00022} _{-0.00022}$  &0.03385 $^{+0.00023} _{-0.00029}$  &0.048 $^{+0.054} _{-0.033}$  &1.510 $^{+0.110} _{-0.084}$ &3.1  $^{+1.1} _{-1.1}$ \\
 c & 3.56004  $^{+0.00012} _{-0.00011}$  &0.04461 $^{+0.00030} _{-0.00038}$  &0.045 $^{+0.051} _{-0.032}$  &2.299 $^{+0.120} _{-0.087}$ &6.3  $^{+1.1} _{-1.2}$ \\
 d & 5.40479  $^{+0.00021} _{-0.00021}$  &0.05893 $^{+0.00040} _{-0.00050}$  &0.043 $^{+0.041} _{-0.030}$  &2.390 $^{+0.104} _{-0.084}$ &7.9  $^{+1.4} _{-1.3}$ \\
 e & 8.26146  $^{+0.00021} _{-0.00022}$  &0.07820 $^{+0.00053} _{-0.00066}$  &0.077 $^{+0.048} _{-0.049}$  &3.390 $^{+0.156} _{-0.110}$ &13.0 $^{+2.0} _{-2.0}$ \\
 f & 12.75758 $^{+0.00050} _{-0.00048}$  &0.10447 $^{+0.00070} _{-0.00088}$  &0.062 $^{+0.064} _{-0.043}$  &2.904 $^{+0.164} _{-0.111}$ & $<$ 8.69 \\
 g & 41.96797 $^{+0.00843} _{-0.00725}$  &0.23109 $^{+0.00154} _{-0.00196}$  &0.059 $^{+0.063} _{-0.040}$  &3.013 $^{+0.303} _{-0.251}$ & $<$ 25.47 \\
\enddata
\end{deluxetable*}

\begin{figure}[t]
\centering
\includegraphics*[scale=.62]{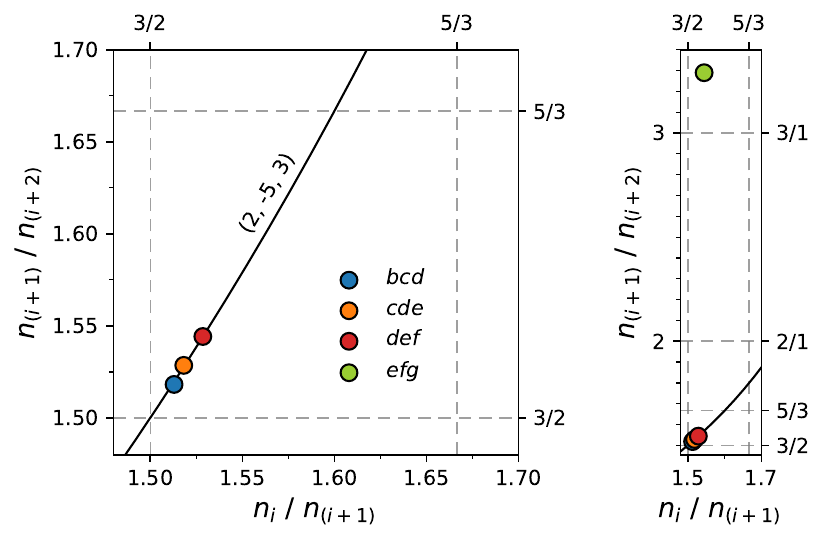}
\caption{Mean-motion ratio plane for K2-138. Colored dots correspond to each set of three adjacent planets 
(triplets). The $x$-axis corresponds to the mean-motion ratio between 
the innermost and middle planet, while the $y$-axis is for middle and 
outermost. Dashed vertical and horizontal lines correspond to 2P-MMRs, while the 
diagonal solid line corresponds to the (2,-5,3) 3P-MMR. Given the large separation 
of $n_f / n_g \sim 3.29$, the $efg$ triplet is shown in the \textbf{right frame}.}
\label{fig:arania}
\end{figure}

As we can see, its architecture is characterised by five tightly packed inner 
planets separated by a significant gap from $m_g$. Relative separations, as 
described by mean-motion ratios between adjacent planets, are $n_b/n_c = 1.513$, 
$n_c/n_d = 1.518$, $n_d/n_e = 1.529$, $n_e/n_f = 1.544$ and $n_f/n_g = 3.290$. 
We use these values in Figure \ref{fig:arania} to locate adjacent planetary 
triplets in the space of mean-motion ratios, where we can study their proximity 
to different resonances (vertical and horizontal dashed lines for 2P-MMRs ; 
diagonal solid lines for 3P-MMRs).

A few different features immediately stand out. The five inner planets are 
located near an intersection of 2P-MMRs, suggesting a 5-planet chain of 
consecutive near first-order 3/2 MMRs and the longest 3/2 chain known to date. 
The green dot, corresponding to the $efg$ triplet, is separated from the rest 
because of the notoriously large gap between $m_f$ and $m_g$. In addition, 
the inner triplets are located on top of the zero-order $(2,-5,3)$ 3P-MMR, 
plotted as the solid black line. Lastly, we note that the offsets which separate 
the planets from the exact 3/2 intersection interestingly increase with the 
distance to the star.

The dynamical history which led K2-138 to its current configuration remains 
uncertain from several angles. For instance: Could the inner planets truly 
be in a 5-planet 3/2 resonance chain in spite of their sizable offsets? The 
origin of the gap between $m_f$ and $m_g$ is also unclear. Why is the sixth 
planet so isolated from the rest given that all underwent a similar migration 
process? Is it really detached from the inner resonance chain?

Both \cite{Lopez.etal.2019} and \cite{Hardegree-ullman.etal.2021} speculate 
about two non-transiting planets at $\sim 19$ and $\sim 30$ days, which would 
complete an 8-planet chain of 3/2 MMRs and help explain the current 
location of $m_g$. However, this is not the only hypothesis.

Considering only the confirmed planets so far, it is not obvious whether $m_f$ 
is detached from the inner chain or not. It is possible that during the migration 
process $m_g$ was captured in a 3/1 MMR with $m_f$, and the large observed 
resonance offset ($\sim$0.3) is either primordial or a consequence of dynamical 
evolution after the dispersal of the protoplanetary disk. In such a case, all 
six planets would have formed a complete resonance chain, at least at some point 
of their history.

While this idea is intriguing, it is unclear how such a large offset may have 
been achieved. There are two main processes which are frequently invoked to 
explain deviations from an observed mean-motion ratio with respect to the 
nominal resonant value: differential migration in a flared-disk 
\citep{Ramos.etal.2017} and tidal effects 
\citep[e.g.][]{Lithwick.Wu.2012,Delisle.Laskar.2014}. Although these effects are 
dependent on the masses, in both scenarios the offset is expected to decrease 
with larger semi-major axis, contrary to what is observed. Moreover, the 
observed mean-motion ratio between the two outermost planets appears 
notoriously large, particularly for a second-order resonance where the family 
of zero-amplitude solutions deviate only slightly with respect to the exact
resonance position. 
  
In this work, we study the cases of both a detached (non-resonant) and an attached (resonant) outermost 
planet $m_g$ and test under which conditions each may have occurred. Section 
\ref{sec:planet_migration} describes the migration framework as well as the 
protoplanetary disk adopted for our N-body simulations of Type-I migration.
Section \ref{sec:mass-regimes} estimates the necessary masses for the 
two outermost planets leading to the detached or attached configurations. In 
Section \ref{sec:5-chain} we discuss the first possibility: a 5-planet resonance 
chain $m_b-m_f$ plus the outermost planet $m_g$ as a non-resonant distant 
companion. Section \ref{sec:6-chain} focuses on the second option and analyses
how very large resonance offsets may be achieved in multi-resonant systems. In 
both scenarios we analyze what would be the necessary planetary masses and disk 
parameters to reproduce the observed dynamical features and how robust these 
solutions are likely to be. Finally, the {\it pantographic effect} leading to increasing offsets is explained in detail in Section \ref{sec:pantograph}, while a 
general discussion and future prospects close the paper in Section 
\ref{sec:conclusions}.

\section{Planetary Migration} \label{sec:planet_migration}

\subsection{Migration Scheme} \label{sec:mig_scheme}

The gravitational interaction between a low-mass planet and a protoplanetary 
disk results in orbital decay, usually known as Type-I planetary migration.
An analytical prescription was originally developed by \cite{Tanaka.etal.2002}, 
who considered a single fully formed planet in a circular orbit and a 
three-dimensional isothermal gaseous disk with a surface density $\Sigma$ and 
aspect ratio $H$ given by
\begin{equation}
\Sigma(r) = \Sigma_0 \left(\frac{r}{r_0}\right)^{-s},  \quad
H(r) = H_0 \left(\frac{r}{r_0}\right)^f,
\label{eq:disc}
\end{equation}
where $s$ and the disk flare index $f$ characterise the surface density 
distribution and shape of the disk, respectively. If $f=0$, the disk profile is 
linear (flat), while $f>0$ describes a curved (flared) disk. Both 
$(\Sigma_0,H_0)$ are the values at $r_0 = 1$ au.

The expression for the orbital decay $\tau_a$ was proposed by 
\cite{Tanaka.etal.2002} for the case described. Later on, 
\cite{Tanaka.Ward.2004} extended the model for planets in eccentric orbits, 
which would circularize when embedded in a disk. The expressions for both the 
migration and circularization timescales are given by
\begin{equation}
\tau_{a} \equiv \left|\frac{a}{\dot{a}}\right| = \frac{Q_a t_{{\rm 
w}}}{H(r)^2} \;\;\; ; \;\;\; 
\tau_{e} \equiv \left|\frac{e}{\dot{e}}\right| = Q_e \frac{t_{{\rm w}}}{0.78}
\label{eq:tau-ae}
\end{equation}

\noindent where $Q_a^{-1}\simeq 2.7+1.1 s$ and we assume the ad-hoc eccentricity 
damping factor adopted by \cite{Cresswell.Nelson.2006} of $Q_e = 0.1$. Lastly, 
$t_{\rm{w}}$ is the wave timescale given by
\begin{equation}
\quad t_{{\rm w}} = \frac{m_0}{m_p} \frac{m_0}{\Sigma(r) r^2} 
\frac{H(r)^4}{\Omega_K(r)}   ,
\label{eq:tau-w}
\end{equation}

\noindent where $m_p$ is the mass of the planet under consideration and $m_0$ 
the mass of the star. The Keplerian angular velocity is given by 
$\Omega_K(r) = \sqrt{G (m_0+m_p)/r^{3}}$. Note that equations (\ref{eq:tau-ae}) 
and (\ref{eq:tau-w}) show that the migration velocity is directly proportional 
to the planetary mass, as well as dependent on the disk parameters. Thus, the 
mass distribution in a given system is largely responsible for the convergence 
or divergence between two or more planets, as well as the possibility of 
resonance trapping \citep[see][]{Beauge.Cerioni.2022}.

In order to account for the coupling between semi-major axis and eccentricities 
during the migration process, \citet{Goldreich.Schlichting.2014} modified 
expressions (\ref{eq:tau-ae}) with a $\beta$ factor to quantify the conservation 
on the angular momentum. The effective timescale for the radial drift is 
\begin{equation}
\frac{1}{\tau_{a_{{\rm eff}}}} = \frac{1}{\tau_{a}} + 2\beta 
\frac{e^2}{\tau_{e}}.
\label{eq:tau-aeff}
\end{equation}
When $\beta=0$ there is no coupling while $\beta=1$ implies no loss in the 
orbital angular momentum. \citet{Goldreich.Schlichting.2014} suggest $\beta = 
0.3$ following estimations by \citet{Tanaka.Ward.2004}.

\subsection{Disk Parameters} \label{sec:disc_parameters}

For our N-body simulations and analytical estimates we assumed a thin laminar 
flat disk ($f=0$) characterized by a surface density exponent of $s=1/2$ and 
an aspect ratio of $H_0 = 0.05$ at $r_0 = 1$ au. We considered several values 
of the surface density $\Sigma_0$ from $100$ up to $1700 \mathrm{\ gr\ 
cm^{-2}}$. In most cases we observed no significant difference in the outcome. 

These parameters were chosen for the sake of simplicity considering our lack of 
information of the real disk. At the very least, we expect our results to be 
robust with respect to reasonable variations of the disk parameters. Equations 
(\ref{eq:tau-ae}) and (\ref{eq:tau-w}) show that $\tau_a$ is directly 
proportional to $H_0$, and inversely so to $s$, $f$, and $\Sigma_0$; we can  
therefore predict the evolution timescale and necessary integration time of our 
simulations as a function of these parameters.

While the general trend of the migrating system is expected to be fairly 
robust to disk parameters, the conditions of resonance capture and the 
subsequent orbital evolution of the resonance chain may be more sensitive. For 
this reason, each new set of parameters led to a new calculation of capture 
conditions with our analytical model and new N-body simulations. More details 
will be given in the next section.

\section{Mass estimates for detachment}\label{sec:mass-regimes}

As may be seen in Table 1, even if fairly reasonable estimations of planetary 
masses are available for the first four planets, only upper bounds are given for 
the two outermost bodies. \cite{Beauge.Cerioni.2022} showed that the formation 
of a resonance chain is particularly sensitive to the mass of the last (and 
first) planet, so the estimation of ($m_f$,$m_g$) can provide detachment regimes 
for the sixth planet.

In that spirit, we start by checking which combination of masses would produce a 
complete 6-planet MMR chain. This calculation does not require N-body 
simulations but may be undertaken using an analytical resonant capture condition 
for $N$ planets developed by \cite{Beauge.Cerioni.2022}. Truth be said, this 
method only checks whether a given (e.g. current) distribution of orbital 
separations is an attractor under the effects of the exterior non-conservative 
force and thus evolves towards more compact configuration, or whether the 
evolution will tend to disperse the system. 

\begin{figure}[t]
\centering
\includegraphics*[scale=.615]{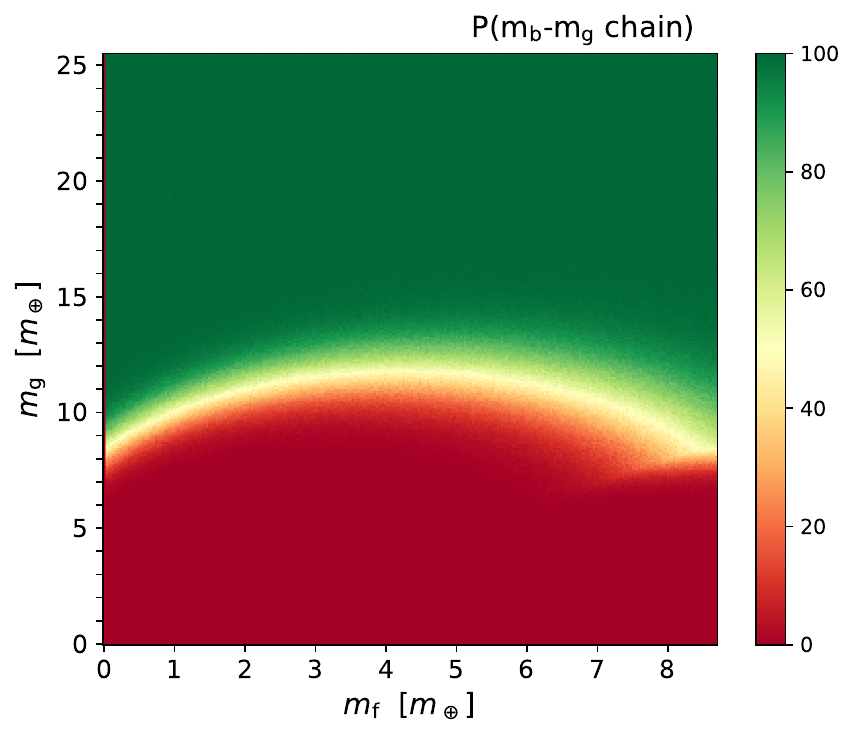}
\caption{Percentage of fictitious 6-planet systems $m_b-m_g$ that 
lead to global convergent migration and the formation of a complete resonance 
chain. While $m_f$ and $m_g$ are varied in a regular grid, for each cell the 
values of the other masses are drawn from a PDF consistent with Table 
\ref{tab:planets}.}
\label{fig3a}
\end{figure}

As mentioned earlier, resonance capture conditions are sensitive to the 
combination of planetary masses, so in Figure \ref{fig3a} we studied a grid of 
values ($m_f$,$m_g$), where each mass varied between zero and its upper bound. 
For each pair of values we generated a set of $N_{\rm sys} = 10^3$ fictitious 
systems varying the mass of the other planets from a probability distribution 
function (PDF) consistent with their observed values\footnote{For each planet 
$i$ with a measured mass of ${m_{i}}_{-\Delta_L}^{+\Delta_R}$, we produced a 
Skew-Normal distribution with median in $m_i$ and such that its integral in the 
interval [$m_i -\Delta_L$,$m_i + \Delta_R$] yields 0.683 (i.e the 
68.3\%-credibility interval) and used this PDF to draw random masses.}. Then, we 
checked what percentage of $N_{sys}$ led to the formation of a stable resonance chain. The color assigned to each cell represents this ratio.

As shown by the color code on the right, the percentage of positive 
values (i.e. convergent infall of the complete 6-planet system) varies from 
zero for low values of $m_g\lesssim 7 \, m_\oplus$, to mixed values around $m_g 
\sim 8-12 \, m_\oplus$, to $100 \%$ for $m_g \gtrsim 13 \, m_\oplus$. For the 
migration of $m_g$ onto the inner system, these three regions respectively 
represent: i) convergent migration leading to an attached $m_g$ and a 6-planet 
chain, ii) a mixed domain of slow divergence/convergence, and iii) divergent 
migration leading to a detached $m_g$ and only an inner 5-planet chain.
These limits are only weakly dependent on the adopted value of $m_f$ and the 
rest of the planets.

We will consider these mass estimates in the following sections in order to 
analyze two options: an $m_g$ planet which is detached from the inner 5-planet 
resonance chain, and $m_g$ planet attached to the inner system and forming a 
complete 6-planet resonance chain.

\section{Option 1: A 5-Planet Resonance Chain}\label{sec:5-chain}

\subsection{A Non-Resonant yet Non-Hierarchical $m_g$}\label{sec:div_mg}

With the innermost planet currently located at $a_b \simeq 0.03$ au around a 
$0.93 m_\odot$ star, it is fair to say that the planets underwent a large-scale 
migration from their original formation sites. 

In this section we assume a non-resonant sixth planet, which means that for 
$m_g$ we are avoiding the green region in Figure \ref{fig3a} and considering 
the red and white areas instead. These are different mass regimes which imply 
different migration rates. We will take it case by case.

To begin with, if the mass of the outer planet were significantly lower than 
$7 \, m_\oplus$ (red tones), then the infall rate of the outermost planet 
would have been much smaller than that of the resonance chain, leading to a 
marked hierarchical planetary system with a much larger relative separation 
than observed. Recall that the mean-motion ratio between the two outer planets 
is $n_f/n_g \simeq 3.3$ and not too far from the 3/1 MMR. 

Therefore, in order to explain the non-resonant but also non-hierarchical 
location of $m_g$, the only option is the white-toned region, corresponding to 
$m_g \sim 8-12 \, m_\oplus$. In this regime, the migration rate of $m_g$ can be 
slightly smaller or higher than that of the inner resonance chain, leading to a 
very slow divergent or convergent migration, depending on the particular 
combination of $m_e, \,m_f$, and $m_g$. The scenario of slow divergence would 
have seen an early formation of the $m_b-m_f$ chain followed by a slow but 
persistent increase in the separation with $m_g$, coming to a halt at the 
observed separation right at the time of dispersal of the gas disk. The same 
would be true for the case of a slow convergent migration of an initially 
farther out $m_g$ planet cut short before it could reach the 3/1 resonance.
While plausible, these alternatives are critically sensitive to the combination of 
initial separation, disk lifetime and differential migration rates.
Nevertheless, given that this section only pretends to study the formation of 
the inner chain with a non-resonant $m_g$, we will take its value from this 
regime.

As a final aside, it is interesting to note that a non-resonant nature for 
$m_g$ may be proposed as strong evidence against an inner disk edge, or, if 
present, one that did not play an important role in the migration history of 
the K2-138 system. If the resonance chain $m_b-m_f$ was stopped at such an edge, 
then $m_g$ would have ultimately reached its inner companions independently of 
its mass. Smaller values of $m_g$ would imply longer times, but the outcome 
would have been the same.

\subsection{Planetary Masses}\label{sec:planet_masses}

\begin{figure}[t]
\centering
\includegraphics*[scale=.615]{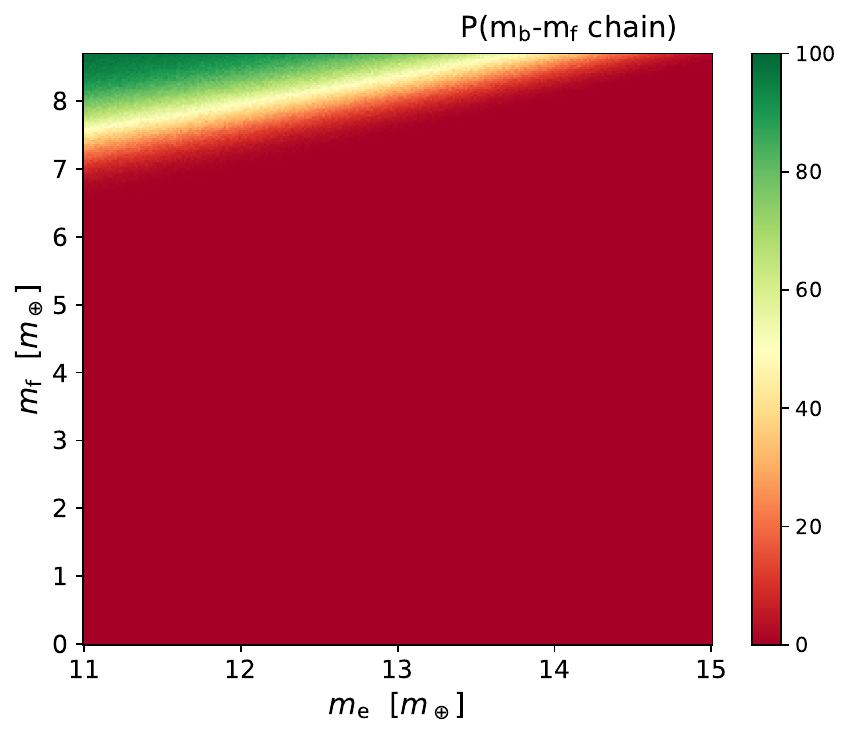}
\caption{Analogous to Figure \ref{fig3a}, but now considering only the 5-planet 
system comprised of $m_b-m_f$. In both plots the color bars indicates the ratio 
of systems which converge to a chain.}
\label{fig3b}
\end{figure}

Assuming that the sixth planet will not come into resonance with the inner 
system, the next question is about the formation of the inner 5-planet chain on 
its own. In Figure \ref{fig3b} we perform a similar analysis to that of Section 
\ref{sec:mass-regimes}, this time focusing on which values of $m_e$ and 
$m_f$ lead to the formation of the inner chain. The values of $m_f$ were varied 
between zero and the upper bound given in Table \ref{tab:planets}, while 
for its inner neighbor we considered values in the range $m_e \in [13 - 2, 13 + 
2] \, m_\oplus$. As before, for each cell of the rectangular grid we generated a 
series of fictitious systems with the masses of the other planets drawn from a 
PDF consistent with the observational data. 

The formation of the 5-planet resonance chain seems very unlikely for the 
estimated parameters of the system. As mentioned earlier, the convergence 
condition is particularly sensitive to the values of the first and last planet 
in the system \citep[][]{Beauge.Cerioni.2022}{}{}. Since the best fit values 
imply $m_f < m_e$ (even for the upper bound of $m_f$), convergent migration 
is highly improbable, especially for $m_e$ larger than its reported median 
value. The color bar shows some positive results, and with a high probability, 
but only for $m_e \sim 11 \, m_\oplus$ and $m_f \simeq 8.5 \, m_\oplus$. While 
the outer planet is still less massive than its inner neighbor, an early 
formation of a resonance chain comprised of $m_b-m_e$ would slow the infall of 
the inner system sufficiently to allow $m_f$ to catch up and join its kin.

\begin{figure*}[t]
\centering
\includegraphics*[scale=.521]{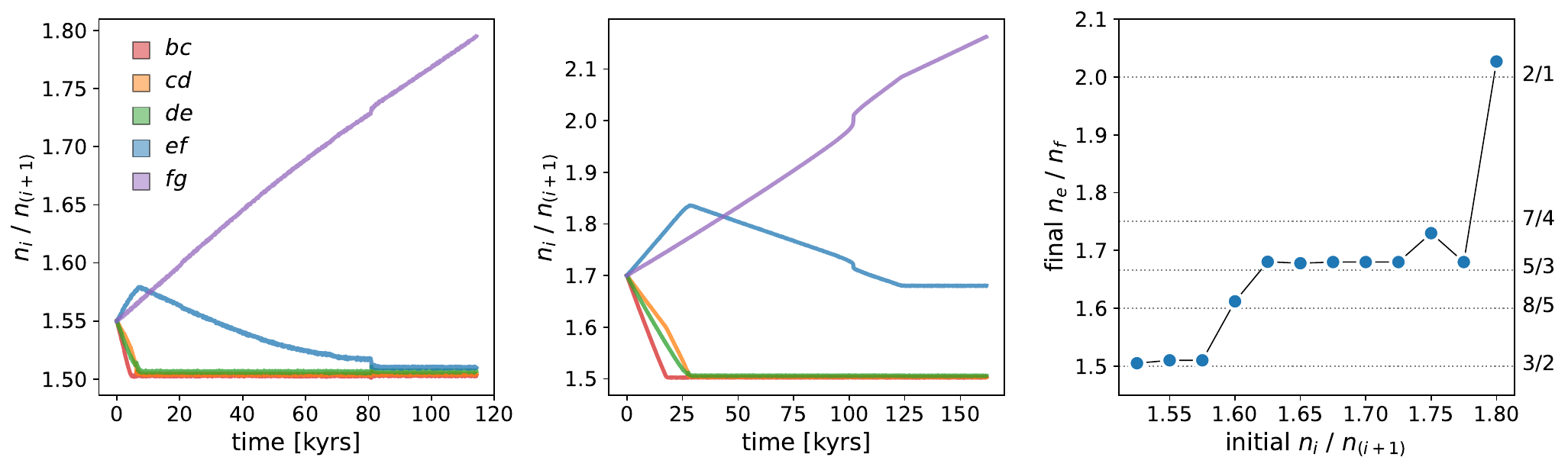}
\caption{\textbf{Left Frame:} N-body simulations under mutual gravitational perturbations 
and migration of the 6-planet K2-138 system. Masses were taken equal to $\{m_k\} 
= (3.1,6.3,7.9,10.0,8.7,8.0) \, m_\oplus$, while for the disk parameters we 
adopted $(s,f,\Sigma_0,H_0) = (0.5,0,500\ {\rm gr/cm}^2,0.05)$. All planets 
were initially placed on circular coplanar orbits with separation such that 
$n_i/n_{(i+1)} = 1.55$. \textbf{Center Frame:} Same as previous plot, but with initial 
separations equal to $n_i/n_{(i+1)} = 1.70$. \textbf{Right Frame:} Final value of 
$n_e/n_f$ as function of the initial separation.}
\label{fig4}
\end{figure*}

\subsection{Initial Separations}\label{sec:init_sep}

A more subtle difficulty in the 6-planet system is related to their initial 
separation leading to the migration process. Even considering the largest value 
for $m_f$ and reasonably low values for $m_e$ (say, $10 \, m_\oplus$), we 
still have mass ratios $m_e/m_f \sim 8.69/10 < 1$. From \cite{Ramos.etal.2017} 
we know that the mass ratio between adjacent planets $m_i$ and $m_{i+1}$ 
leading to 2-planet convergent migration must satisfy the condition
\be
\frac{m_{i+1}}{m_i} > \left( \frac{n_i}{n_{i+1}} \right)^{\frac{2}{3}(2f+s-1/2)}
\ee
where $n_i/n_{i+1}$ is not the resonant mean-motion ratio we wish to achieve 
but the initial value at the start of the migration. A flat disk ($f=0$) and a 
surface density exponent $s=1/2$ give a minimum value, i.e. $m_{i+1}/m_i = 1$ 
independent of the initial separation. Other disk parameters, especially those 
with a steeper surface density, lead to more stringent conditions for the mass 
ratio. 

Since the estimated value of $m_e/m_f$ is smaller than unity, the inner body of 
the pair migrates faster and $m_f$ lags behind increasing any initial 
separation. Convergent migration between these two planets is only expected to 
occur after the inner resonance chain $m_b-m_e$ is established and the effective 
semi-major axis decay rate is slowed. This behavior is expected for all disk 
properties and regardless of the initial conditions, although the resulting MMR 
in which each pair is captured may (and will) vary.

Nevertheless, there is one hard condition on initial separations. Given that 
the mass ratios of the first four planets is greater than unity, migration is 
convergent and they will be trapped in a mean-motion commensurability that is lower 
than their initial separations. Therefore, we set initial separations greater 
than $n_i/n_{(i+1)} > 1.5$.

\subsection{N-Body Simulations}\label{sec:simus}

We performed N-body simulations employing a Bulirsch-Stoer integration scheme with variable time-step and precision $ll=12$ \citep{Bulirsch1966}. 
All bodies begin with initial separations $n_i/n_{(i+1)} = 1.55$ and migration rates as outlined in \ref{sec:mig_scheme}.
The left frame of Figure \ref{fig4} shows results for the 
aforementioned scenario, where we plot the mean-motion ratio of each adjacent 
pair of planets as a function of time.  

The mass ratio increases subsequently for the first three pairs, so migration is 
convergent and they rapidly are trapped in a four-planet chain of MMRs. However, 
since $m_f < m_e$ the fifth planet starts the simulation falling behind and 
increasing the value of $n_e/n_f$. This process reverses once the inner system 
is trapped. The net infall speed of the four-planet chain is now lower than the 
decay rate of $m_f$ \citep[see][]{Beauge.Cerioni.2022} and this planet 
ultimately catches up and joins the resonance chain. No such fate awaits the 
outermost planet and $m_g$ continues to fall behind increasing its distance with 
the rest of the pack.

The scenario described above appears consistent with the overall dynamical 
characteristics of the system, and we were able to obtain similar results for a 
wide range of disk properties. Variations in the masses also led to analogous 
results as long as the mass ratios satisfied the convergence/divergence 
conditions mentioned earlier and outlined in Figures \ref{fig3a} and 
\ref{fig3b}. Different results, however, were found when changing the initial 
separation between the planets. The central frame of Figure \ref{fig4} shows a similar 
N-body simulation as before, but now considering a wider initial system with 
$n_i/n_{(i+1)} = 1.70$. 

Since the planets start the simulation further apart, the migration of the 
inner pack towards the 3/2 MMRs takes longer, giving time for $m_f$ to separate 
even more and encounter a larger number of commensurabilities in its path. When 
the differential migration reverses, $m_f$ does not reach the 3/2 MMR with 
$m_e$ but becomes trapped in the 5/3, leading to a configuration inconsistent 
with the one observed. The right frame shows how the final value of the 
ratio $n_e/n_f$ varies according to the initial separation between adjacent 
planets. 

Unsurprisingly, initial separations close to the 3/2 MMR yield final systems 
trapped in such a commensurability and thus analogous to the observed K2-138 
6-planet model. However, even slightly wider separations lead to other 
resonances and other configurations, with the 5/3 and 2/1 being the most 
common, depending on the initial $n_i/n_{(i+1)}$. Although extremely high 
migration rates will sometimes avoid all these alternate capture sites and fall 
back to the 3/2, this was rarely the case. The capture condition for a given 
resonance depends much more strongly on the mass ratio than on the surface 
density of the disk, and planets with similar masses will suffer similar 
semi-major axis decay rates even for high values of $\Sigma_0$. We found results 
analogous to Figure \ref{fig4} even for $\Sigma_0 = 1700$ gr/cm$^2$, but we did 
not test higher values. 

These simulations seem to indicate that the current 3/2 MMR resonance chain 
among the five inner planets of K2-138 is possible provided the mass of $m_f$ 
is close to its upper bound, the mass of $m_e$ is significantly lower than its 
median value (beyond the 1 or 2 $\sigma$-level), and the initial separation 
between the planets was such that their mean-motion ratios all lied between 
$1.5$ and $1.6$. This would require that the planets form around the same time and with very similar mean-motion ratios as those observed today. While all these conditions are, again, possible, they seem to 
require very fine tuning and some of the parameters (such as the mass ratio 
$m_e/m_f$) are very far from the expected values. 

\begin{figure}[t]
\centering
\includegraphics*[scale=.73]{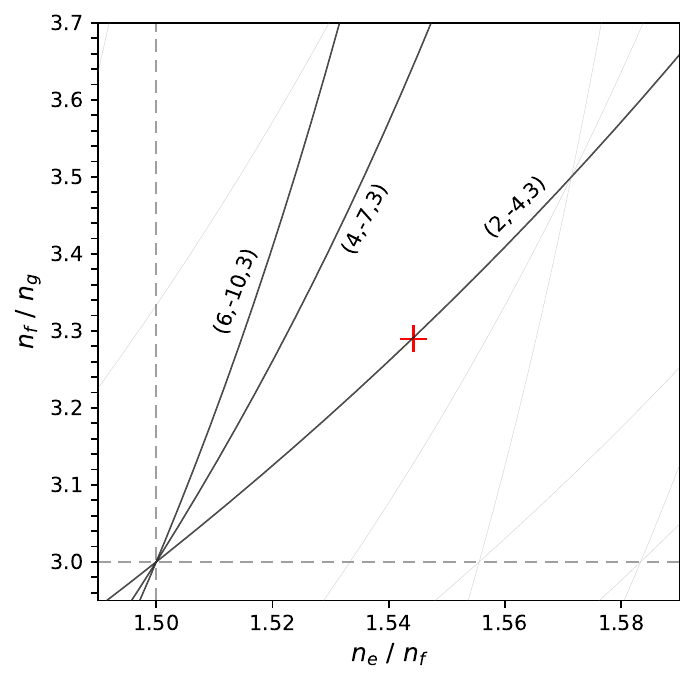}
\caption{Observed position of the $(n_e/n_f,n_f/n_g)$ triplet (red cross) in 
the mean-motion ratio space. Diagonal curves mark every low-integer zero- and 
first-order 3P-MMR that passes through the domain. Commensurabilities crossing 
the $(3/1,3/2)$ intersection are labeled and drawn using thicker lines. Error bars are too small to be distinguishable in the figure.}
\label{fig:acordeon}
\end{figure}

\section{Option 2: A 6-Planet Resonance Chain}\label{sec:6-chain}

\subsection{The (2,-4,3) 3-Planet Resonance}\label{sec:2-4-3}

The idea that the outermost planet ($m_g$) may form part of the system's 
resonance chain is strongly suggested when we analyze the location of the 
triplet $(n_e/n_f,n_f/n_g)$ in the mean-motion ratio plane. This is shown with a 
red cross in Figure \ref{fig:acordeon}, together with the position of every 
zero- and first-order $(k_1,k_2,k_3)$ 3P-MMR in the region, with indexes 
$|k_i|\le10$. The 2-planet commensurabilities $n_e/n_f = 3/2$ and 
$n_f/n_g = 3/1$ are drawn in dashed lines, while 3P-MMRs are shown in 
continuous lines. Of these, three resonances cross the $(3/1,3/2)$ intersection 
and are highlighted with thicker lines and labeled in the plot.

The location of the three outer planets in this plane is strikingly close to the 
$(2,-4,3)$ first-order three-planet resonance, characterized by the relation 
$2n_e - 4n_f + 3n_g = 0$. Introducing the observed values for the orbital 
periods from Table \ref{tab:planets}, as well as their uncertainties, we obtain 
\be
2n_e - 4n_f + 3n_g = 0.00020^{+0.00024}_{-0.00015}\ d^{-1}
\label{eq6}
\ee

which comes extremely close to the exact nominal commensurability. Given the limited number of two and three-planets resonances in 
this region, it seems very unlikely that such a correlation could be a 
coincidence, even if the 3P-MMR is of first order and thus not the strongest 
dynamical feature in the vicinity. Similarly to the inner triplets, Figure 
\ref{fig:acordeon} suggests a capture of $(n_e/n_f , n_f/n_g)$ in the $(3/2 , 
3/1)$ intersection, with an excursion of the triplet over the three-planet 
resonance that crosses that intersection.

To prove that the $(2,-4,3)$ resonance is not only relevant to the current 
dynamics of the system but also tied to its migration history, we must solve two 
independent issues: (i) test whether planetary masses consistent with the outer 
K2-138 system can be trapped in this first-order 3P-MMR and survive the 
dissipation of the gaseous disk without being expelled from the 
commensurability, and (ii) explain the large separation of the current 
configuration with respect to the $(3/2,3/1)$ intersection of two-planet 
resonances.

\subsection{Formation of the 6-Planet Chain}\label{sec:capture}

In principle, there are two possible routes that could lead to the capture of 
the outer triplet in the $(2,-4,3)$ resonance; these are divergent and 
convergent migration. 

Divergent migration occurs in a scenario similar to that discussed in the 
previous section, but with a mass of the outer planet in the range $m_g \lesssim 
8 m_\oplus$ (Figure \ref{fig3a}) and initial separations such that $n_f/n_g < 
3$. In the previous section we assumed all planets initially equidistant forming 
a compact formation; this is not strictly necessary and the outer planet could 
have began its migration further out. However, we could not find any set of 
parameters that led to the capture in the $(2,-4,3)$ three-planet 
commensurability during divergent migration, so this case is ruled out. As a 
side note, some cases did result in the zero-order $(4,-7,3)$ 3P-MMR, with 
stable orbits and a small but non-zero offset with respect to the two-planet 
resonances.

We then turn to the second possibility, a convergent migration of $m_g$ from an 
initial orbit beyond the 3/1 resonance with $m_f$ in the hope of trapping the 
three outermost planets in the desired first-order 3P-MMR. As before we fix the 
masses of the four inner planets to the values chosen in the previous sections: 
$(m_b,\ldots,m_e) = (3.1,6.3,7.9,10.0) m_\oplus$ and with initial separations 
among adjacent bodies between 1.5 and 1.6. The rest of the masses, as well as 
their initial separations, and disk parameters were varied. 

\begin{figure}
\centering
\resizebox{1\columnwidth}{!}{\includegraphics{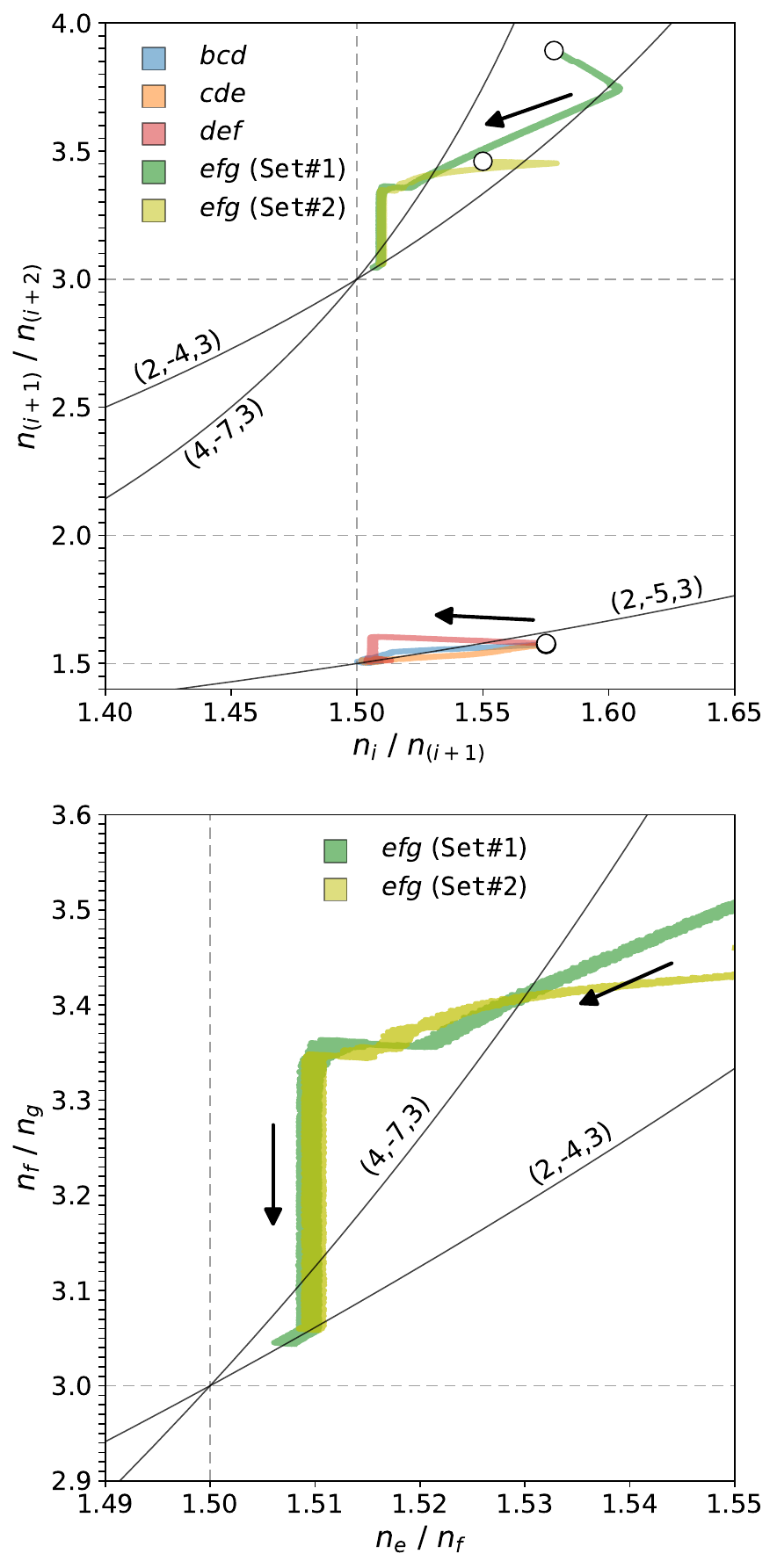}}
\caption{
N-body simulations under mutual gravitational perturbations and migration of the K2-138 system in the mean-motion ratio plane.
Colored lines correspond to each planetary triplet. Inner triplets belong to {\tt Set\#1}, but analogous results are obtained for {\tt Set\#2}. The evolution of of the $efg$ triplet in both sets are shown in green colors. Initial positions are shown with white circles. Arrows indicate the direction of evolution. \textbf{Bottom frame} is a zoom-in of the resting place of the last triplet.}
\label{fig:243}
\end{figure}

\begin{figure*}
\centering
\resizebox{2.12\columnwidth}{!}{\includegraphics{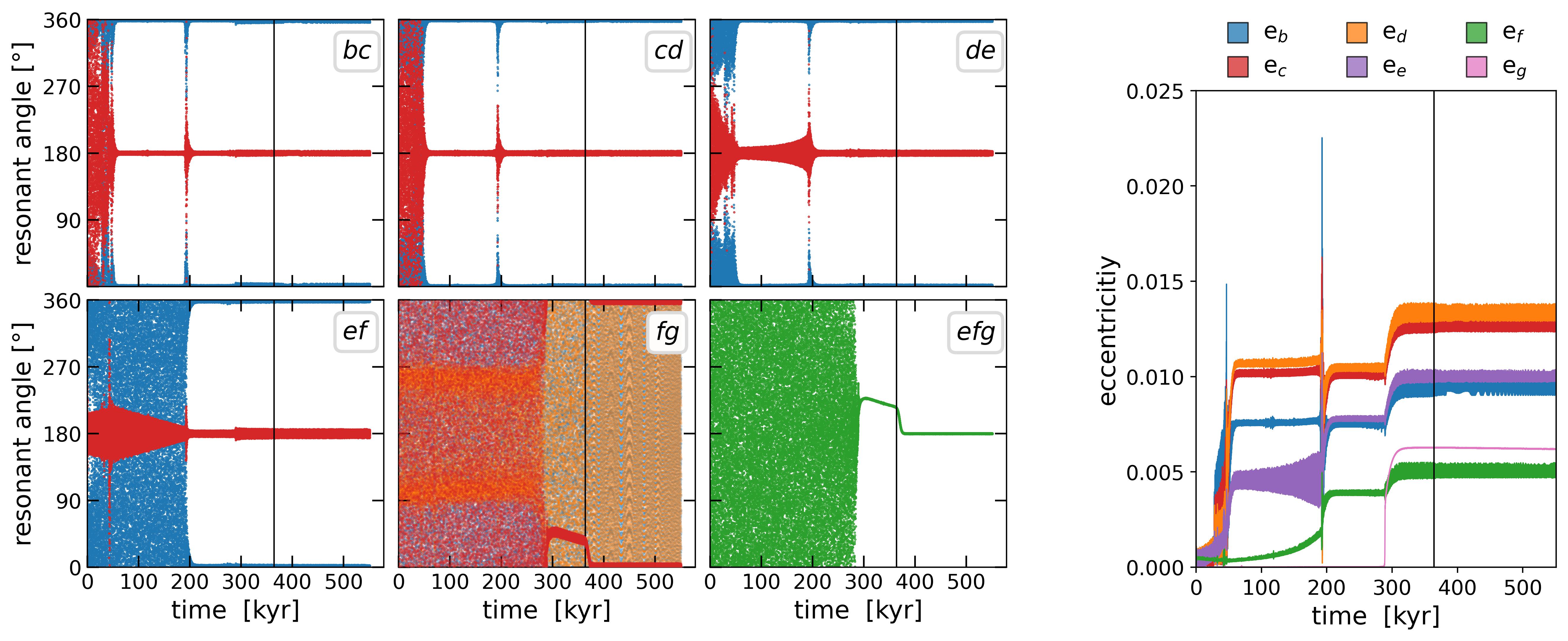}}
\caption{\textbf{Left Frame:} Time evolution of resonant angles corresponding to the simulation {\tt Set\#1}. The first four plots on the left show the resonant angles $\sigma_1 = 3\lambda_{i+1} - 2 \lambda_i - \varpi_i$ (in blue) and 
$\sigma_2 = 3\lambda_{i+1} - 2 \lambda_i - \varpi_{i+1}$ (in red) for the first 
four inner planetary pairs (i.e. $i = 1,2,3,4$).
The lower center plot shows the resonant angles of the 3/1 MMR between $m_f$ and $m_g$, where blue is 
used for $\theta_1 = 3\lambda_g - \lambda_f - 2\varpi_f$, red for $\theta_2 = 
3\lambda_g - \lambda_f - \varpi_f - \varpi_g$ and orange for $\theta_3 = 
3\lambda_g - \lambda_f - 2\varpi_g$. The capture of the pair into the two-planet 
resonance is only accompanied by a libration of $\theta_2$ with the other angles 
circulating.
The green data points in the lower-right frame show the time 
evolution of the resonant angle $\theta_{243} = 2\lambda_e - 4 \lambda_f + 
3\lambda_g - \varpi_g$ corresponding to the $(2,-4,3)$ 3P-MMR of the outer 
triplet. \textbf{Right Frame:} Eccentricities of all planets for the same simulation. The disk is smoothly removed from the simulation at $T = 375 $ 
kyrs (vertical black line). All subsequent orbital variations occur in a 
gas-free scenario.}
\label{fig:captura}
\end{figure*}

Figure \ref{fig:243} shows the results of two different simulations, labeled 
{\tt Set\#1} and {\tt Set\#2}. Their main characteristics are given below:
\be
\begin{split}
{\rm Set\#1 : } & \;\; m_f = 9.3 \;,\; m_g = 11.0 \;,\; \Sigma_0 = 100 \; {\rm 
g/cm^2} \\ 
{\rm Set\#2 : } & \;\; m_f = 8.7 \;,\; m_g = 8.8 \;,\; \Sigma_0 = 1000 \; {\rm 
g/cm^2}\\ 
\end{split}
\label{eq7}
\ee
where the masses are given in units of $m_\oplus$. 
In {\tt Set\#1} we tested a value of $m_f$ that is slightly over the reported 
upper bound as low values of this parameter prevent the formation of the inner 
chain of multiple 3/2 resonances (Figure \ref{fig3b}). Moreover, the mass 
of the outer planet in {\tt Set\#2} was set barely large enough to guarantee 
convergent migration once the inner system was trapped in its chain. This means 
a very slow convergent migration w.r.t its inner neighbors and thus its 
initial semi-major axis could be placed close to the 3/1 MMR with $m_f$; we 
chose $n_f/n_g \simeq 3.4$, although slightly smaller values lead to similar 
results. Conversely, the mass of $m_g$ in {\tt Set\#1} was significantly larger 
leading to a convergent migration of the outer planet onto $m_f$ even before the 
resonance lock of the inner system. In consequence the initial separation had to 
be larger indicative of a more hierarchical primordial system. We opted for 
$n_f/n_g \simeq 3.8$, although larger values would result in similar final 
configurations.  

The top frame of Figure \ref{fig:243} shows the evolutionary tracks of the 
mean-motion ratios of all triplets, each identified with a different color. The 
large circles indicate the initial values, and arrows show the direction of the 
evolution. The behavior of the inner system $m_b-m_f$ was similar in both 
sets, resulting in the formation of the well-known 3/2 resonance chain. The 
evolution of the outer triplet is shown in green, where a dark tone was chosen 
for {\tt Set\#1}, and a lighter color for {\tt Set\#2}. 

Both simulations show fairly similar characteristics. Since $m_f < m_e$, 
initially $n_e/n_f$ increases signaling a divergent migration of this pair. As 
soon as the inner resonance chain begins to assemble, their migration rate slows 
and allows $m_f$ to reverse its differential migration. The evolutionary track 
thus switches direction and the pair $m_e-m_f$ moves towards the 3/2 MMR where 
it is finally trapped. The behavior of the outer pair is simpler as the value 
of $m_g$ guarantees convergent migration in both cases. However, while the 
mass chosen for {\tt Set\#1} leads to a relatively fast inward migration from 
the start of the run, in {\tt Set\#2} the outer pair approaches the rest of the 
system only after the resonance chain begins to form. This explains the almost 
horizontal evolutionary track during the first part of the simulation.

The bottom frame of Figure \ref{fig:243} zooms in on the vicinity of the final 
resting place of the outer triplet. After the capture of $n_e/n_f$ in the 
3/2 MMR both runs are practically indistinguishable. As the mean-motion 
ratio of the outer pair $n_f/n_g$ approaches the 3/1 MMR, it first passes 
through the $(4,-7,3)$ 3P-MMR but is finally stopped at the $(2,-4,3)$ 
commensurability. Even though this first-order resonance should have a lower 
capture probability than its near neighbor, we found it surprisingly robust and 
most of our simulations led to the same result. 

Figure \ref{fig:captura} shows the time evolution of several resonant angles 
and eccentricities for {\tt Set\#1}. {The first four plots on the left frame show $\sigma_1 = 3\lambda_{i+1} - 2 \lambda_i - \varpi_i$ (in blue) and 
$\sigma_2 = 3\lambda_{i+1} - 2 \lambda_i - \varpi_{i+1}$ (in red) for the first 
four inner planetary pairs (i.e. $i = 1,2,3,4$).
The inlaid box in the upper-right corners of each frame shows the planets involved in each case. All 
four pairs evolve towards an ACR \citep[see, 
e.g.,][]{Beauge.etal.2003,Michtchenko.etal.2006} characterized by equilibrium 
values $(\sigma_1,\sigma_2) = (0,\pi)$. As each link of the resonance chain 
falls into place the resonant angles and eccentricities (right frame in the figure) suffer a temporary kick, but are rapidly restored to low-amplitude 
oscillations around their stationary solutions.

The lower center plot shows the resonant angles of the 3/1 MMR between $m_f$ 
and $m_g$. The capture of the pair into the two-planet resonance is only 
accompanied by a libration of $\theta_2 = 3\lambda_g - \lambda_f - \varpi_f - 
\varpi_g$ while all other possible resonant angles circulate. At the same 
time the outer pair enters the 3/1 commensurability, the outer triplet is 
captured in the $(2,-4,3)$ 3P-MMR. The green data points in the lower-right plot 
show the time evolution of the resonant angle $\theta_{243} = 2\lambda_e - 4 
\lambda_f + 3\lambda_g - \varpi_g$ displaying an asymmetric libration during the 
time interval up to $\sim 375$ kyrs. All other possible resonant angles 
circulate. 

It is difficult to prove that the capture in the three-planet resonance is a 
consequence of the trapping of the individual pairs in the 2-planet 
commensurabilities; the fact that the system chose the first-order 3P-MMR over 
its zero-order neighbor is intriguing; additional work is necessary before 
establishing in detail how this stable configuration is reached. 

In order to test the stability of such a multi-resonant configuration, we 
introduced a smooth reduction in the surface density of the disk during a time 
interval centered around $375$ kyrs and which lasted about ten thousand years. 
The subsequent evolution of the system occurs without the effects of disk 
damping. The librations of both the 3/1 MMR and the three-body resonance 
angles turn towards symmetric solutions as the surface density of the disk 
tends to zero, with $\theta_{2} \rightarrow 0$ and $\theta_{243} \rightarrow 
\pi$. The orbital evolution of the full 6-planet resonance chain remains stable 
throughout the simulation with no indication of any incipient increase in 
amplitudes that could signal a future disruption of the system.

\subsection{Tidal Evolution of the 6-Planet Chain}\label{sec:tidal}

While we have succeeded in trapping the outermost triplet in the $(2,-4,3)$ 
3P-MMR and thus constructing a full 6-planet resonance chain, the offsets with 
respect to the nominal 2-planet resonances are much lower than observed. This 
is particularly notorious in the case of $n_f/n_g$ whose current value ($\simeq 
3.29$) is so distant from the exact resonance that it does not appear related.

The rest of the planetary pairs are also significantly displaced from the exact 
3/2 resonance, with observed mean-motion ratios equal to $1.513$, $1.518$, 
$1.529$ and $1.544$, respectively. None of our capture simulations were able to 
obtain similar values, regardless of the chosen planetary masses or disk 
parameters. Consequently, if our previous results are representative of the 
primordial system, the current resonant offsets must have originated after the 
dissipation of the primordial disk; tidal effects with the central star appears 
as the only possible mechanism.

Several studies have analyzed how tidally-induced orbital variation generated in 
an inner planet may propagate to outer companions where tidal direct effects 
with the central star are negligible. In the case of secular (non-resonant) 
systems, this synergy affects mainly the eccentricities, although the finite 
equilibrium eccentricity induced by the mutual gravitational interactions also
affects the rate of semi-major axis decay 
\citep[e.g.,][]{Rodriguez.etal.2011,Greenberg.VanLaerhoven.2011}. The case of 
resonant configurations, particularly involving 3-planet systems, is more 
intricate. Even though tidal evolution generates divergent migration and, thus, 
escape from 2-planet resonances, \cite{Papaloizou.2015} found that in the case 
3 planets in a double resonance the escape occurs along a 3P-MMR 
commensurability. Similar results were reported by \cite{Charalambous.etal.2018} 
under a wide range of initial conditions and migration rates.

\begin{figure}
\centering
\resizebox{1.0\columnwidth}{!}{\includegraphics{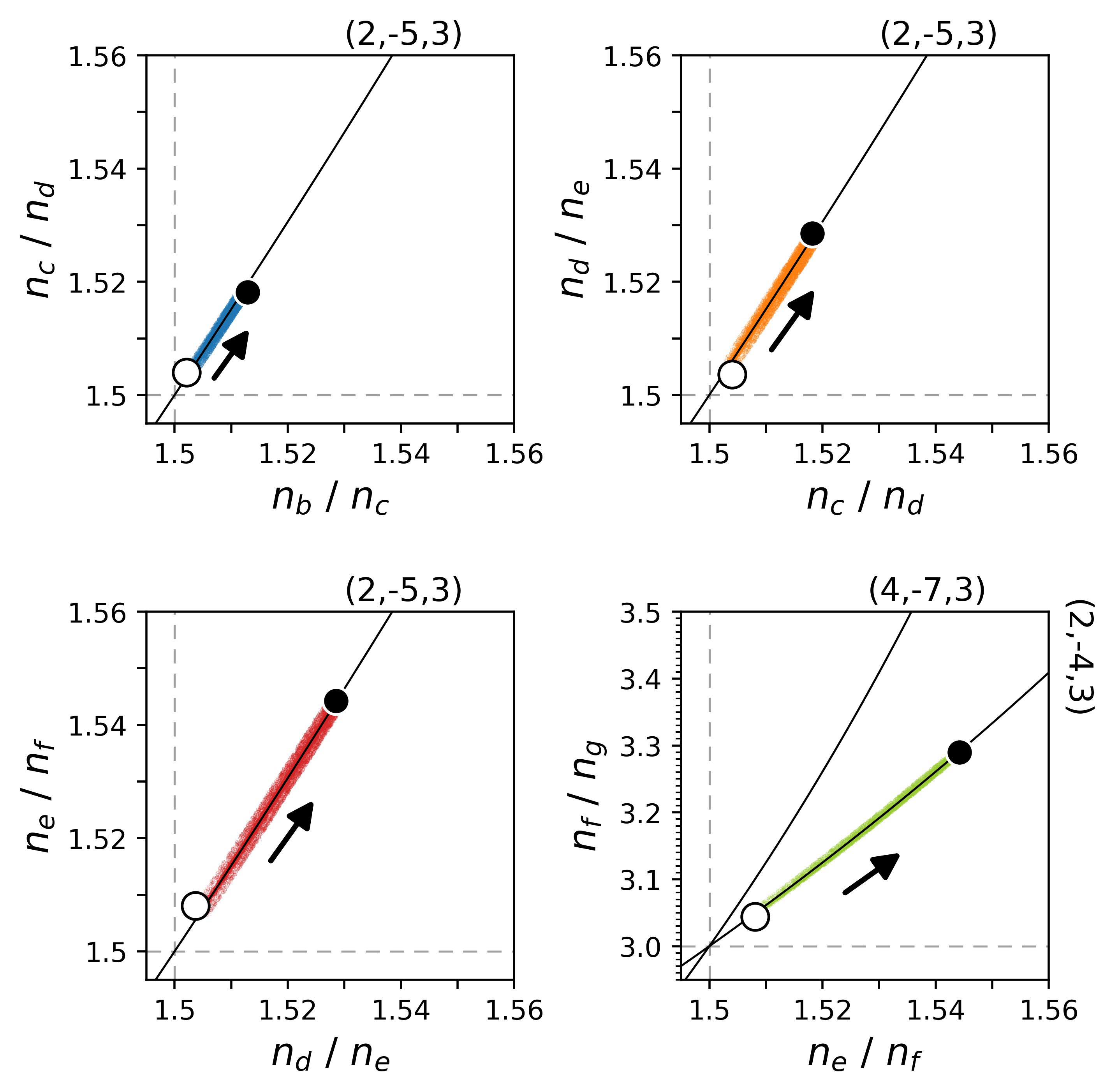}}
\caption{Tidal evolution of the K2-138 system in the mean-motion ratio plane. Colored lines correspond to the triplets of {\tt Set\#1}. Initial positions (white circles) 
correspond to the final stage of the capture simulation discussed in the 
previous section, with semi-major axes scaled to the observed system. The 
current mean-motion ratios of the planets are shown in black circles. Analogous results are obtained for {\tt Set\#2}.}
\label{fig:tidal}
\end{figure}

\begin{figure}
\centering
\resizebox{1.04\columnwidth}{!}{\includegraphics{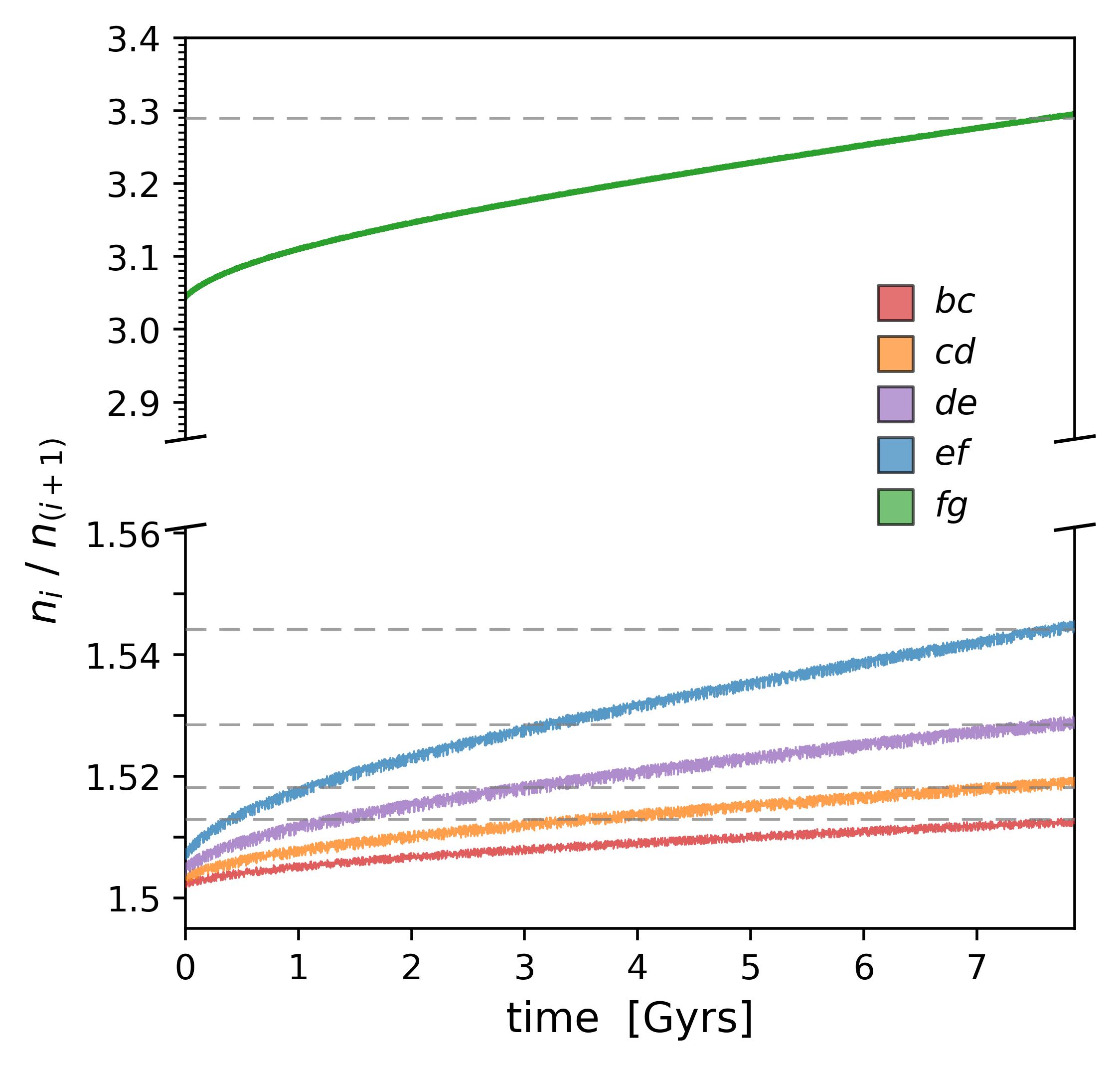}}
\caption{Time evolution of the mean-motion ratios of adjacent planets due 
to tidal interactions with the central star. As with the previous plot, initial 
conditions correspond to the final stage of the capture simulation of {\tt 
Set\#1}. Observed values are highlighted with horizontal dashed lines.}
\label{fig:tidal_set1}
\end{figure}

The offset observed in several planetary systems with respect to 2P-MMRs (e.g. 
Kepler-80, TOI-178) appear consistent with such an evolutionary route 
\citep[e.g.,][]{MacDonald.etal.2016, Leleu.etal.2021}, perhaps indicating that 
the resonant structure of small-mass planetary systems could have a stronger 
correlation to 3P-MMRs than to two-planet commensurabilities 
\citep{Cerioni.etal.2022}. However, in no case was a capture reported in a 
first-order three-planet resonance, nor was such a configuration suggested as 
an evolutionary route due to tidal effects. 

In order to check such a hypothesis for K2-138, we performed a series of N-body 
simulations of {\tt Set\#1} and {\tt Set\#2} under the combined effects of 
mutual gravitational perturbations and tides. The initial conditions were taken 
equal to the end product of the capture simulations, scaling the semi-major axes 
to better fit the observed system. Eccentricities and angles were untouched. 

The tidal evolution was simulated following 
\citep{Mignard.1979}. This model considers tidal bulges raised on the planets as well as on the central star, which will be slightly delayed from (or ahead of) their instantaneous equipotential positions by a constant time-lag (CTL).
We assumed a stellar tidal parameter $Q'_0 = 10^6$, while for the 
planets we chose $Q'_i = 10^2$, regardless of the body's mass or initial orbit. 
Recall that in the CTL model these numerical values are defined at the 
onset of the simulation and change during the orbital evolution of the system  
such that $Q'_i n_i = const$. 

Figure \ref{fig:tidal} shows results for {\tt Set\#1}; analogous outcomes were
obtained for {\tt Set\#2}. Each plot shows the evolutionary tracks of 
mean-motion ratios associated to adjacent planetary triplets. Initial values are 
depicted with white circles while the evolution of the system in these planes 
is shown with colored lines; arrows are drawn to help visualize the direction of 
the evolution. As expected, tidal effects increase the 2-planet resonance 
offset while preserving each triplet in its three-planet commensurability. This 
effect is also noted in the case of the outer triplet where the guiding route 
is the first-order 3P-MMR characterized by the index array $(2,-4,3)$. 
Finally, the current mean-motion ratios of the planets are identified with black circles.

While not explicitly shown, semi-major axes decrease due to the tidal 
interactions, and this effect is strongest for the innermost planet. This 
differential migration pulls open the first planetary pair, which in turn opens 
up every other pair in the resonance chain, much like the motion of a 
pantograph-like mechanism. This is what is driving the evolution shown 
in Figure \ref{fig:tidal}. More details will be given in Section 
\ref{sec:pantograph}. We can thus see that the observed values can in fact be 
obtained with tidal evolution, and that this mechanism is effective even in the 
case of the outer pair $n_f/n_g$.

Additional proof is given in Figure \ref{fig:tidal_set1} which shows the 
evolution of each mean-motion ratio over time; current values are highlighted 
with horizontal dashed lines. Not only is the observed orbital separation 
consistent with the tidal evolution of the system, but the individual offsets of 
all planetary pairs are attained at the same time. 

In these simulations, and for 
the adopted values of the tidal parameters, the current state of the system was 
reached at approximately $8$ Gyrs after the dissipation of the primordial disk. 
Stellar age estimated from spectral analysis yields $T_{\rm age} = 
2.3^{+0.44}_{-0.36}$ Gyrs \citep{Lopez.etal.2019}. This requires a faster tidal evolution and thus a smaller 
value of $Q'_i$ for the planets; we find that tidal parameters of the order of $Q'_i \simeq 
30$ would have led to the same configuration in integration times consistent 
with $T_{\rm age}$. However, stellar ages are notoriously difficult to evaluate 
with precision, and the actual value for this system could be larger. 

\begin{figure}[t]
\centering
\resizebox{1.03\columnwidth}{!}{\includegraphics{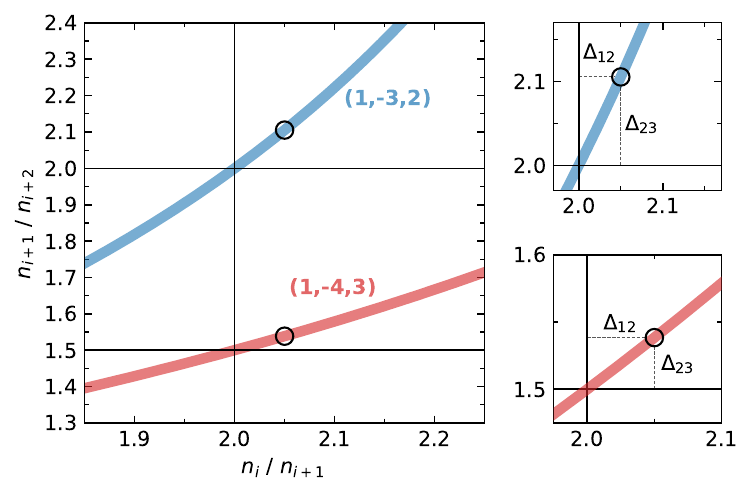}}
\caption{Sketch of two 3P-MMRs in mean-motion ratio space. Points correspond to 
triplets locked in the respective three-body resonances. \textbf{Right frames} are 
zoomed in with equal-scale axes, in order to better compare the slope of each 
curve, which determines the ordering of the two-body resonant offsets: 
$\Delta_{12}<\Delta_{23}$ for the blue curve and $\Delta_{12}>\Delta_{23}$ for 
the red.}
\label{fig:acor}
\end{figure}

\section{The Pantographic Effect}\label{sec:pantograph}

We repeated the previous simulations ignoring tidal interactions between the 
star and all but the innermost planet ($m_b$); results were indistinguishable, 
indicating that any change in the resonant offsets is mainly driven by the 
gravitational interactions with the other bodies and not by direct tidal forces. 
Thus, the current offsets of K2-138 may be used to estimate the tidal parameter 
of the inner planet but not of the rest of the system. 

The observed increase of the resonance offsets with the semi-major axis is not a 
consequence of tidal interactions but of interplay between planets in a 
resonance chain. In order to understand the reason behind this effect, 
let us concentrate on a generic triplet $m_1$, $m_2$, and $m_3$ trapped in a 
double resonance but slightly away from the nominal commensurabilities:
\be
\frac{n_1}{n_2} = \frac{p_1+q_1}{p_1} + \Delta_{12}
\hspace*{0.4cm}  ;  \hspace*{0.4cm}
\frac{n_2}{n_3} = \frac{p_2+q_2}{p_2} + \Delta_{23}
\label{eq8}
\ee
where $\Delta_{12}$ and $\Delta_{23}$ are the offsets initially determined by 
the capture process. Additionally, we assume that the three planets are also
trapped in a $(k_1,k_2,k_3)$ three-body commensurability with no noticeable 
offset, such that

\be
k_1 n_1 + k_2 n_2 + k_3 n_3 = 0 .
\label{eq9}
\ee

\noindent where the order of the 3P-MMR is given by the sum of indexes $k_1+k_2+k_3$. 
Dividing equation (\ref{eq9}) by $n_2$ and replacing equation (\ref{eq8}), we 
get a condition for the offsets $\Delta_{12}$ and $\Delta_{23}$ that must be 
satisfied in the face of 2P- and 3P-MMRs:
\be
\Delta_{23} = -k_3\ (k_2\ +\ k_1\ \frac{p_1+q_1}{p_1}\ +\ k_1\ \Delta_{12})^{-1}\ -\ \frac{p_2 + q_2}{p_2} .
\label{eq10}
\ee
Expression \ref{eq10} shows that the offset of the outer pair, 
apart from being precisely determined by the first, can also be either smaller 
or higher than it, depending on the resonances involved. The same result can be deduced 
from analyzing 3P-MMR slopes in the mean-motion ratio plane, as the offsets will 
be the triplet's coordinates with the origin placed at the 2P-MMR intersection. 

Figure \ref{fig:acor} shows two examples. A triplet trapped in the $(1,-4,3)$ 
commensurability close to the $n_1/n_2 = 2/1$ and $n_2/n_3 = 3/2$ will have 
$\partial (n_2/n_3)/\partial (n_1/n_2) < 1$ and thus the offset of the outer 
pair will be smaller than that of the inner pair. The opposite behavior is 
observed in the case of the three planets trapped in the $(1,-3,2)$ 3P-MMR. In 
such a configuration the outer pair will suffer a larger departure from the 
nominal two-planet resonance even if $m_1$ is the only body driving the 
migration process.  

\begin{figure}[t]
\centering
\resizebox{1.015\columnwidth}{!}{\includegraphics{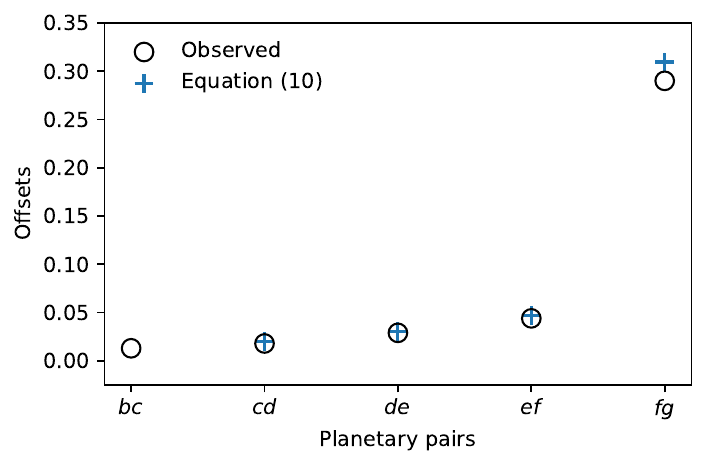}}
\caption{Open black circles plot the observed two-planet 
resonance offsets for the K2-138 system. Blue crosses show the predictions given 
by expression (\ref{eq10}) assuming the current $\Delta_{bc}$ as a starting 
value.}
\label{fig:offsets}
\end{figure}

The same principle can be applied to any number of interacting triplets in a 
long resonance chain, such as K2-138, where the $(2,-5,3)$ and $(2,-4,3)$ 
3P-MMRs have slopes $> 1$ and the offsets increase with semi-major axis.
Furthermore, Expression \ref{eq10} shows that only one given offset will 
determine the rest, and that if any planetary pair is spread apart or brought 
together, every other pair will simultaneously follow, while the triplets travel 
over the curves of 3-planet resonances. This linked motion holds a resemblance 
to the crossing points in a \textit{pantograph}-like mechanism, and is what 
happens in Figure \ref{fig:tidal} when tidal effects bring the innermost planet 
inwards, pulling open the first planetary pair.

Figure \ref{fig:offsets} shows, in blue crosses, the observed offsets for each 
planetary pair. As mentioned previously, the values grow with the distance to 
the star with a surprisingly large jump in the case of $\Delta_{fg}$ due to the 
pronounced slope of the (2,-4,3) resonance. The open black circles show the 
values predicted from Equation ({\ref{eq10}) starting from the observed value of 
$\Delta_{bc}$. The agreement with the real values is excellent in the case of 
the first three links, while some discrepancy is noted in the case of the outer 
pair. This difference is due to the accumulation of short-period deviations in 
each successive pair of planets.

\section{Discussion}\label{sec:conclusions}

In this paper we analyzed the current dynamical configuration of the complete 
K2-138 system. While the resonance chain involving the five innermost planets 
($m_b$ up to $m_f$) is an expected outcome of planetary migration, the orbital characteristics of the outer 
planet $m_g$ are intriguing. Its large distance from the rest and apparent non-resonant dynamics strike out in such an otherwise compact system.

We analyzed two possible solutions to explain the dynamics of $m_g$. In the 
first, a small-mass outermost planet could have migrated slower than the inner 
system and become detached as the planets migrated closer to the star. This 
requires a small mass for $m_g$ as well as fine-tuned initial separations and 
disk parameters. While possible, this formation scenario appears unlikely. 

A second, and more interesting, explanation arose when noting that the three 
outer planets seem very close to the $(2,-4,3)$ first-order 3P-MMR. At first 
glance this proximity did not appear relevant given that these resonances are 
generally believed to be weak. Moreover, the mean-motion ratio $n_f/n_g \simeq 
3.3$ appears too far from any significant 2-planet commensurability, thus 
making any correlation between the current separation and a past resonance 
capture very unlikely. A closer look, however, showed both beliefs unfounded. A 
more massive outer planet with $m_g \gtrsim 9 m_\oplus$ initially located 
beyond the 3/1 MMR with $m_f$ could be trapped in a double resonance $(n_e/n_f\ ,\ n_f/n_g) = (3/2\ ,\ 3/1)$ leading to 
a libration of the critical angle of the 3P-MMR $(2,-4,3)$. This outcome was found 
for a wide range of initial conditions and disk parameters. The resulting 
configuration at the time of the dissipation of the primordial disk was 
composed of all 6 planets in a stable resonance chain with small offsets.

Tidal evolution of the system for times comparable with the system's age proved 
sufficient to convert the small resonance offsets to their present-day values, 
at least for values of $Q'$ of the order of $Q'_b \simeq 30$ for the innermost 
planet. Direct tidal interactions between the other planets and the star proved 
negligible and thus their tidal parameters could not be constrained. Surprisingly even the current 
offset of the outer pair $n_f/n_g \simeq 3.3$ can be explained by tidal 
evolution alone, indicating that perhaps other seemingly non-resonant systems 
may be dynamically affected by two-planet or three-planet commensurabilities,
even more so than previously thought \citep[e.g.,][]{Cerioni.etal.2022}.

The propagation of resonance offsets among members of a resonance chain can 
take several forms, depending on the 2P-MMRs and 3P-MMRs involved in each link. 
As a result, offset-inducing mechanisms such as tidal effects will not produce offsets that necessarily decrease with the distance from the star, as previously thought.
Offsets among chained planets will also be linked such that the opening (contraction) of any single pair will simultaneously spread apart (bring together) the rest, a behavior
reminiscent of the extension/contraction of a pantograph. 

While at present 
it is not possible to prove that K2-138 forms a complete 6-planet resonant 
chain, including a first-order 3P-MMR, we believe the dynamical evidence is 
sufficient for such a claim to be at least very plausible.

\acknowledgments
The authors would like to express their gratitude to the computing facilities 
of the Instituto de Astronomía Teórica y Experimental (IATE) and the 
Universidad Nacional de Córdoba (UNC). This work was funded by research grants 
from the Consejo Nacional de Investicaciones Científicas y Técnicas (CONICET) 
and the Secretaría de Ciencia y Tecnología (SECYT/UNC).

\bibliography{k2-138}{}
\bibliographystyle{aasjournal}

\end{document}